\documentclass{emulateapj}                  
\usepackage{graphicx,amsmath}

\slugcomment{Accepted ApJ June 2005}
\begin{document} 
\shorttitle{Spiders in Lyot Coronagraphs}
\shortauthors{Sivaramakrishnan \& Lloyd}
\newcommand \cf     {cf.}
\newcommand \eg     {{\it e.g., }}
\newcommand \etc    {{\it etc}}
\newcommand \etal   {{\it et~al.}}
\newcommand \ie     {{\it i.e.,}}
\newcommand \viz    {{\it viz.,}}
\newcommand \lap    {$\stackrel{<}{\sim}$}
\newcommand \gap    {$\stackrel{>}{\sim}$}
\newcommand \fwhm   {${FWHM}$}
\newcommand \both   {(\Phi + {\Psi}_k)}
\newcommand \pk     {{\Psi}_k}
\newcommand \ph     {\Phi}
\newcommand \Sumk   {\Sigma_k\ }
\newcommand \eq     {\,=\,}                 
\newcommand \sgn   {{\rm{sgn}}}
\newcommand \sinc   {{\rm{sinc}}}
\newcommand \jinc   {{\rm{jinc}}}
\newcommand\conv {{ * }}

\title{Spiders in Lyot Coronagraphs}

\author{Anand Sivaramakrishnan, \altaffilmark{1} }
	\affil{Space Telescope Science Institute\\
		3700 San Martin Drive, Baltimore, MD 21218}
		\and
\author{James P. Lloyd \altaffilmark{1}}
	\affil{Department of Astronomy\\
		230 Space Sciences Building\\
		Cornell University\\
		Ithaca, NY 14853}
\altaffiltext{1}{NSF Center for Adaptive Optics}
\begin{abstract} 
	In principle, suppression of on-axis stellar light by a coronagraph is
	easier on an unobscured aperture telescope than on one with an
	obscured aperture.  Recent designs
	such as the apodized pupil Lyot coronagraph, the `band-limited' Lyot
	coronagraph, and several variants of phase mask coronagraphs work best
	on unobscured circular aperture telescopes.  These designs are developed to
	enable the
	discovery and characterization of nearby jovian or even terrestrial exoplanets.
	All of today's major space-based and adaptive optics-equipped ground based
	telescopes are obscured aperture systems, with a secondary mirror held in
	place by secondary support `spider' vanes.  The presence of a secondary
	obscuration can be dealt with by ingenious coronagraph designs, but the spider
	vanes themselves cause diffracted light that can hamper the search for jovian
	exoplanets around nearby stars.  We look at the problem of suppressing spider vane
	diffraction in Lyot coronagraphs, including apodized pupil and band-limited
	designs.  We show how spider vane diffraction can be reduced drastically, and
	in fact contained, in the final coronagraphic image, within one resolution element
	of the geometric image of the focal plane mask's occulting spot.
	This makes adaptive optics coronagraphic searches for exojupiters possible
	on the next generation of adaptive optics systems 
	being developed for 8--10~m class telescopes such as  Gemini and the Very Large Telescopes.
\end{abstract}

\section{Introduction}

In the last few years several exciting new coronagraphic designs
have been proposed.  These designs focus on suppressing starlight within a
few resolution elements of bright stars in order to detect planetary
companions of these stars fractions of an arcsecond from the star,
a task which requires a contrast ratio of more than a million to one 
for extrasolar jovian planets \citep{BSH04, Bnat05}.
An even more ambitious goal, that of detecting and characterizing extra-solar
terrestrial planets, demands contrast ratios of a billion to one
or higher.

Recent progress in coronagraphic concepts has yielded several classes of coronagraphs
that achieve the requisite suppression in the ideal case 
\citep[{\it e.g.},][]{KT02, SAF03,SDA03,Kasdin03,Guyon03}, 
which generally requires zero wavefront error, and unobscured apertures.  
The fully-optimized, diffraction-limited Lyot project coronagraph
\citep{Lyot39, Sivaramakrishnan01, OSM03, OppenheimerSPIE04, DigbySPIE04, MSP05},
which is deployed at the Air Force AEOS 3.6m telescope \citep{LRCN02} is the first
coronagraph to operate 
in the regime of  `extreme adaptive optics' (ExAO) with Strehl ratios around 90\% under
the best seeing conditions in the $H$-band.   This instrument, and all future
ground and space coronagraphs, will have to work with non-zero 
wavefront error, and on aperture geometries subject to real-world
engineering constraints.  In \citet{Lloyd05} and \citet{Sivaramakrishnan05blc} 
we analyzed the effects of wavefront errors on Lyot coronagraphs.   
In this paper, we focus on the effects of spider diffraction in a Lyot coronagraph.  
 
Next generation AO systems, designed for Gemini and an ESO Very Large Telescope,
are likely to be dedicated to coronagraphic imaging and spectroscopy in the search
for exo-jupiters. It is therefore important to understand and quantify the effects spiders 
have on coronagraphic imaging as these systems are designed and constructed 
over the next five years. This is particularly relevant in light of recent work on high 
dynamic range coronagraphy on obscured apertures \citep{Soummer05}.

We focus on an analytical treatment of this problem with the goal of producing
some general understanding of the problem, not just a calculation for a specific
case at a specific telescope.  Over and above the narrow application of specific 
numerical calculations, there is a computational difficulty in modelling coronagraphs 
on obstructed apertures when taking spider vanes into account. For example, 
coronagraphic image simulations require about 6 to 8 samples across each 
resolution element in order to accurately model the effects of a $4 \lambda/D$ 
image plane occulting stop (`focal plane mask') in the focal plane ($\lambda$ 
being the central wavelength of the observing bandpass, and $D$ the telescope 
diameter).  This entails optical calculations using arrays 6 to 8 times the diameter 
of the entrance pupil.  The 8\ m Gemini telescope's spider vanes are 1\ cm thick.
Thus the numerical arrays required to model the optical train can be
12000--16000 elements on a side.  Furthermore, these numerical calculations
can require extra-ordinary resolution and dynamic range to converge on the 
correct answer, and the effects of aliasing in an FFT can be quite severe at the 
$10^{-9}$ contrast level [\eg\ even very high resolution numerical calculations of 
the Four Quadrant Phase Mask coronagraph \citep{RRB00} are inadequate
to correctly calculate the coronagraphic point-spread function (PSF)],
requiring an analytical solution \citep{LGG03}.  Fourier transform routines of
optical modelling programs will need more memory than can be put in
most of today's computers. We are therefore forced to fall back on developing 
analytical estimates of the effect of secondary mirror support vanes on 
coronagraphic images.  We treat classical Lyot coronagraphs \citep{Lyot39} 
with their hard-edged, opaque focal plane masks; modified Lyot coronagraphs 
with graded focal plane masks described by Gaussian\footnote{Ftaclas, private 
communication} or `band-limited' \citep{KT02} functions; as well as apodized 
pupil Lyot coronagraphs \citep{SAF03, Soummer05}. \citet{Guyon03} 
discusses pure pupil apodized high dynamic range imaging on arbitrarily-shaped 
apertures, \citet{SDA03} treats dual-zone phase-mask coronagraphy on arbitrary 
apertures, and \citet{LGG03} presents ways of dealing with four-quadrant
phase mask coronagraphy on centrally-obscured apertures with spiders.

\section{Monochromatic coronagraphic theory with spider supports}

Here we briefly recapitulate our basic monochromatic Fourier optics formalism.
A more detailed treatment can be found in \citet{Sivaramakrishnan01, Lloyd05}.
We recollect that a plane monochromatic wave travelling in the $z$ direction in 
a homogenous medium without loss of energy can be characterized by a complex
amplitude $E$ representing the transverse (\eg\ electric) field strength of the wave.
The full spatio-temporal expression for the field strength is $E e^{(i\kappa z - 
\omega t)}$, where $\omega/\kappa  = c$, the speed of the wave. We do not use 
the term {\it field} to denote image planes --- the traditional optics usage --- we 
always use the term to denote electromagnetic fields or scalar simplifications of 
them. The wavelength of the wave is $\lambda = 2\pi/\kappa $. The time-averaged 
intensity of a wave at a point is proportional to $EE^*$, where the average is taken 
over one period,  $T = 2\pi/\omega$, of the harmonic wave. The phase of the 
complex number $E$ represents a phase difference from the reference phase 
associated with the wave.   A real, positive $E$ corresponds to an electric field 
oscillating in phase with our reference wave.  A purely imaginary positive value 
of $E$ indicates that the electric field lags by a quarter cycle from the reference 
travelling wave. Transmission through passive, linear filters such as apertures, 
apodizers, and so forth, is represented by multiplying the field strength by the
transmission of these objects which modify the wave.  Again, such multiplicative
modification is accomplished using complex numbers to represent 
phase changes forced on the wave incident on such objects.

We assume that Fourier optics describes our imaging system: image field
strengths are given by the Fourier transform of aperture (or pupil --- we use the 
two terms interchangeably) illumination functions, and vice versa.

\begin{figure*}[htbp]
\epsscale{0.5}
\plotone{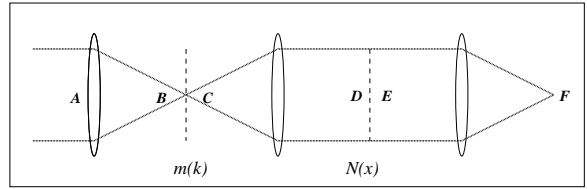}
\caption{The essential planes and stops in a coronagraph.  The entrance aperture
is A, the direct image at B falls on a focal plane mask whose transmission function
is $m(k)$. The re-imaged pupil plane D, after being modified by passage through
a Lyot stop with a transmission function $N(x)$, is sent to the coronagraphic
image at F.} 
\label{fig:corolayout}
\end{figure*}

\begin{figure*}[htbp]
\epsscale{1.0}
\plottwo{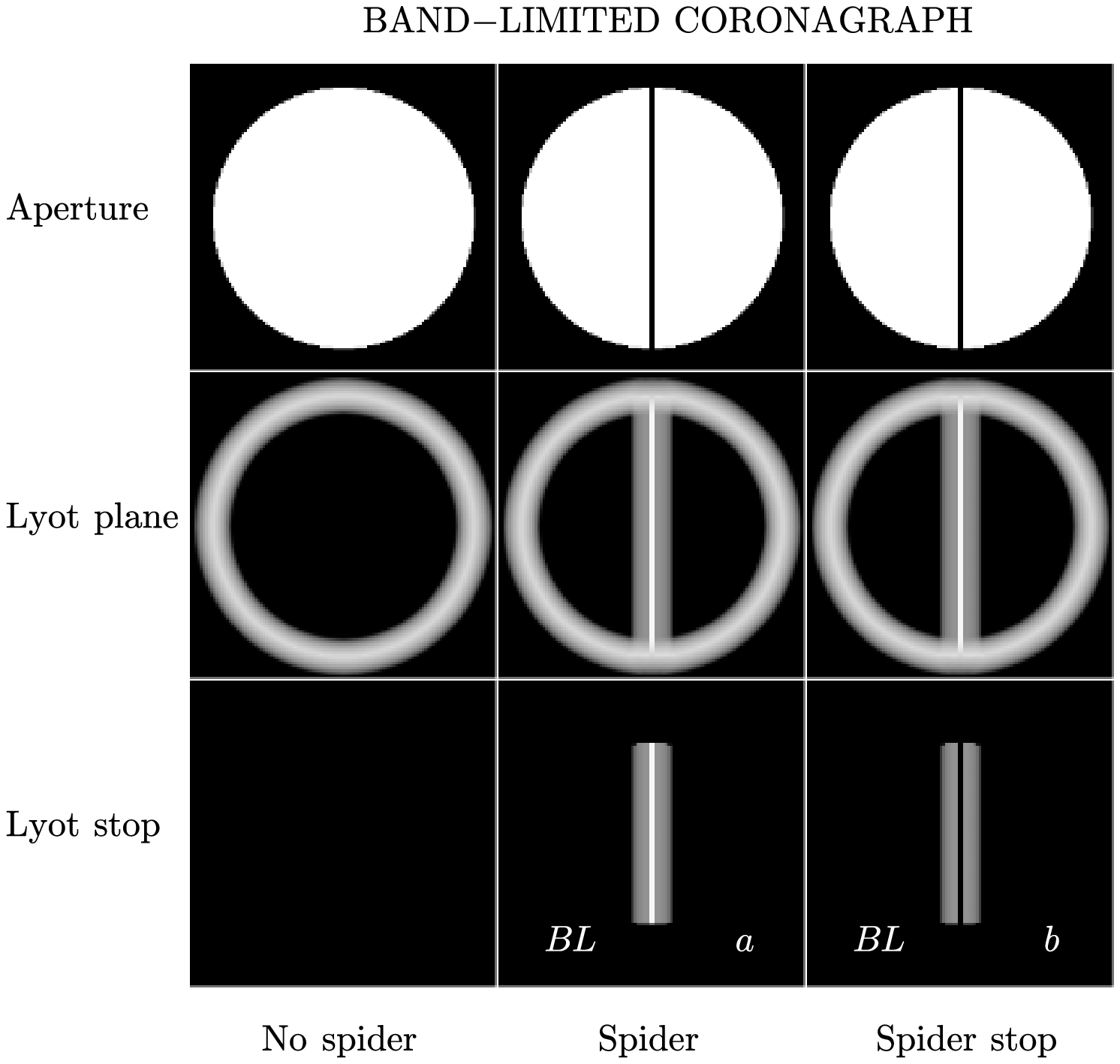} {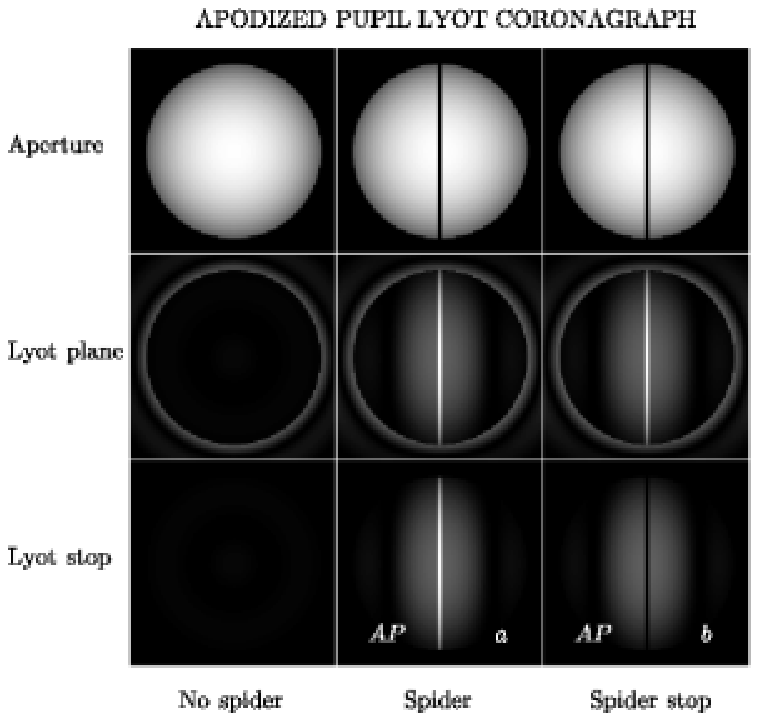}
\caption{Key planes of a monochromatic, band-limited coronagraph and an apodized 
pupil Lyot coronagraph on a circular, unaberrated aperture with and without a single 
spider vane. Intensity in the various pupil planes are shown on the same logarithmic 
greyscale between 1 and $1.0 \times 10^{-4}$. The perfect coronagraph on a clear 
aperture with no spider vane obstruction (``No spider" columns) produces a sufficiently 
dark Lyot pupil {\it interior}, as is seen in the ``Lyot plane" rows. The ``Lyot stop" rows 
show two styles of Lyot stops used. Left and middle, a simple Lyot stop removes all 
incoming on-axis light on the external edge of the entrance pupil. In the presence of a 
spider vane the simple Lyot stop leaves the bright strip of light in the geometrical 
image of the spider vane, and the `aura' of surrounding light, whose intensity is 
proportional to the square of the ratio of the spider vane thickness to the telescope 
diameter. At right, a Lyot stop which removes the bright strip of light located at the 
spider vane position itself.}
\label{fig:pupils}
\end{figure*}

A telescope aperture is described by a transmission function pattern $A(x)$, 
where $x=(x_1,x_2)$ is the location in the aperture, in units of the wavelength of 
the light (see Fig.~\ref{fig:corolayout}). The corresponding aperture illumination 
describing the electric field strength in the pupil (in response to an unaberrated,
unit field strength, monochromatic incident wave)  is $ E_A =  A(x) $. From this 
point onwards we drop the common factor $E e^{(i\kappa z - \omega t)}$ when 
describing fields.  The aperture intensities ($E_A E_A^*$) for two coronagraphic 
designs are shown in Fig.~2 (top row). The field strength in the image plane, 
$E_B = a(k)$, is the Fourier transform of $E_A$, where $k=(k_1,k_2)$ is the 
image plane coordinate in radians. Because of the Fourier relationship between 
pupil and image fields, $k$ is also a spatial frequency vector for a given 
wavelength of light. We refer to this complex-valued field $a$ as the 
`amplitude-spread function' (ASF), by analogy with the PSF of an optical 
system. The PSF is $aa^*$. Our convention is to change the case of a function 
to indicate its Fourier transform.
\placefigure{fig:corolayout}
We multiply  the image field $E_B$ by a mask function $m(k)$ to model the 
focal plane mask of a coronagraph. The image field immediately after this mask is 
$ E_C = m(k) \,  E_B $. The electric field in the re-imaged pupil after the focal plane 
mask (the Lyot pupil, see  Fig.~\ref{fig:pupils}, middle rows) is $E_D$, which is the 
Fourier transform of $E_C$.  We use the fact that the transform of the image plane 
field $E_B$ is just the aperture illumination function $E_A$ itself, so the Lyot pupil 
field is $ E_D =  M(x) \conv  E_A $, where  $\conv$ is the convolution operator.

\placefigure{fig:pupils}
If the Lyot pupil stop transmission is $N(x)$, the electric field after the Lyot stop 
is $E_E = N(x) E_D$.  The transform of this last expression is the final 
coronagraphic image field strength: $ E_F  = n(k) \conv [m(k) \, E_B]$.
In this paper we look at the effects of secondary support spiders on the final 
coronagraphic PSF corresponding to the field strength $ E_F $.  We take into 
account the fact that $A(x)$ may be apodized (Fig.~\ref{fig:pupils}, right), so 
$A$ is a graded function rather than a function that takes values of either 0 or~1.
Understanding high dynamic range Lyot coronagraphy hinges on understanding 
the structure of the field strength $E_D$ in the Lyot plane located at D, as well 
as the repercussions of such structure in the final image plane $ E_F $.

\section{Spider diffraction in a Lyot coronagraph} \label{LyotSpider}

The mask function in a Lyot coronagraph is best expressed as
$m(k) = 1 - w(k)$, where $w(k)$ is the `focal plane mask shape'
function.  For a hard-edged mask $w(k) = \Pi(D|k|/s)$, where $s$
is the mask diameter in units of the resolution of the optical system, 
$\lambda/D$. The function $\Pi(\alpha)$ takes the value unity for 
$|\alpha| < 0.5$, and is zero elsewhere. We note that  $w(0,0) = 1$ 
(which constrains $W(x)$ to have unit area). The Fourier transform  
of the focal plane mask transmission function $m(k)$ is $M(x) = 
\delta(x) - W(x)$, so the Lyot pupil electric field of a Lyot coronagraph
can be expressed as 
\begin{eqnarray} \label{EDL} 
E_D &=&  [\delta(x) - W(x)]  \conv  A(x).
\end{eqnarray}
(An example of the mask transmission function for a simple band-limited 
coronagraph is shown in Fig.~\ref{fig:blcstop}.)

We must understand the morphology of the Lyot pupil field in order to 
understand the extent to which secondary spider supports reduce the 
final coronagraphic image's dynamic range. 

We model a single spider vane across the entrance pupil of a telescope 
by writing 
\begin{eqnarray} \label{EAspider} 
E_A &=&  A(x) ( 1 - \Pi(x_1/\epsilon))
\end{eqnarray}
($\epsilon$ being the width of the spider vane). Secondary obstructions 
result in different effects. We do not treat them here
\citep[see {\it e.g.},][for details on this topic]{Sivaramakrishnan01,LGG03}.
Equation (\ref{EDL}) with this aperture function produces a Lyot pupil field 
\begin{eqnarray} \label{ELspider} 
E_L &=&  A(x)   ( 1  - \Pi(x_1/\epsilon) \conv \delta(x_2))  \nonumber \\
& & -  W(x)) \conv  [A(x) ( 1  
 - \Pi(x_1/\epsilon) \conv \delta(x_2))].
\end{eqnarray}
We are only concerned with the `interior' of the aperture in the Lyot plane, where,
by design, we are satisfied with coronagraphic performance of our aperture without
spider support vanes.  This means that in our estimation, $A(x) - W \conv A(x)$
is sufficiently small for our scientific purposes 
\citep[{\it e.g.},][]{SAF03,Soummer05} or zero \citep{RRB00, Aime01aa, Aime02aa, 
KT02, SDA03}. We therefore drop this component of equation (\ref{ELspider}),
so in the interior of the Lyot pupil we obtain
\begin{eqnarray} \label{ELIspider} 
E_{L,\ {\rm interior}} 
&=&    A(x) [ \Pi(x_1/\epsilon) \conv \delta(x_2) ] \nonumber \\
& & - W(x) \conv  ( A(x) \Pi(x_1/\epsilon) \conv \delta(x_2)).
\end{eqnarray}
This is the contribution (in the Lyot pupil) due to diffraction from a long thin
obstruction such as a spider vane in the entrance aperture.

Fig.~\ref{fig:pupils} (left) shows the intensity of the Lyot pupil field using a
perfect theoretical coronagraph, the band-limited coronagraph, where, in the 
absence of spider vanes and optical aberrations, the interior of the Lyot field 
is identically zero.

There are two components to the field strength in the interior region of the Lyot 
pupil. The bright central stripe is exactly the spider vane width $\epsilon$ --- its 
brightness is very close to that of the brightness of the clear or apodized aperture 
in the entrance pupil. This is the first term in equation (\ref{ELIspider}), \viz\ 
$ A(x) [\Pi(x/\epsilon) \conv \delta(x_2)]$.  For high dynamic range applications it 
can be masked out with a thin opaque strip in the Lyot pupil stop. This strip can 
be oversized for practical reasons without noticeably affecting throughput at the 
Lyot stop.  We call this kind of Lyot stop a `Lyot spider stop'.  It resembles the 
`reticulated Lyot stop' of \citet{SY05}, which masks out bright inter-segment gaps 
in the coronagraphic Lyot plane of extremely large, highly-segmented telescopes.

The extended low-intensity `aura' of the bright spider vane is described by the 
second term  in equation (\ref{ELIspider}), $- W(x) \conv  ( A(x) \Pi(x_1/\epsilon) 
\conv \delta(x_2))$ (Fig.~\ref{fig:pupils}, bottom row). For the applications we 
consider here, the equivalent width of the function $W(x)$ is of order $D/s$, since 
we wish to search for faint companions and structure outside an inner working 
angle of about $ s\lambda/2D$ of a bright, on-axis star.  Typically $s$ will lie 
between 4 and 10.

When the width of the focal plane mask becomes larger than $\sim 10 \lambda/D$,
this strip of dimmer light can be be removed by a Lyot pupil stop with an oversized 
spider vane obstruction without sacrificing Lyot stop throughput too much. We point 
out that the exact geometry of the spiders is not relevant --- our approach can deal
with non-orthogonal spiders as easily as with perfectly aligned spiders.

\subsection{The coronagraphic PSF with spiders}

We now estimate the field strength in the final coronagraphic image with a Lyot 
spider stop. The coronagraphic ASF without a Lyot stop of any kind is the Fourier 
transform of the field in the interior of the Lyot pupil described by equation 
(\ref{ELIspider}):
\begin{eqnarray} \label{ASFspiderstop} 
a_c(k) &=& \epsilon a(k) \conv \sinc(\epsilon k_1) \nonumber \\
& &   - \epsilon w(k)  \bigl{(} a(k) \conv \sinc(\epsilon k_1 )\bigr{)}.
\end{eqnarray}
The first term of the ASF $a_c(k)$ is $\epsilon a(k) \conv [\sinc(\epsilon k_1)]$. 
This has the shape of a regular spider diffraction spike, but is down a factor 
$\epsilon^2$ in intensity from the direct image's spider spike.  Masking out the 
bright spider vane removes this first term in the expression  for $a_c(k)$.  The 
remaining term is modulated by the mask shape function $m(k)$ itself.  If the 
mask shape is the circular top hat function $\Pi(D|k|/s)$, spider diffraction in 
the coronagraphic image plane is confined to the region behind the mask 
when a Lyot spider stop is used. The $\sinc(\epsilon k_1 )$ function 
describes the direct image's spider profile along the length of the spider 
`spike' in the image plane. This function is close to unity in the region behind 
or just around the  focal plane mask in the direct or coronagraphic image 
planes.  $w(k)$ is also unity at scales where $a(k)$ is significant in size if the 
focal plane mask is a few to several resolution elements wide. Thus,  where 
$w(k)$ is non-zero (for hard-edged masks), $a(k) \conv \sinc(\epsilon k_1 )$
is very close to the integral of $a(k)$ over its entire domain. The value of this 
integral is $A(0)$ because of the Fourier relation between $a$ and $A$. We 
stress that we are trying to estimate the size of the spider effects here rather 
than calculate them exactly.

\begin{figure*}[htbp]
\epsscale{0.5}
\plotone{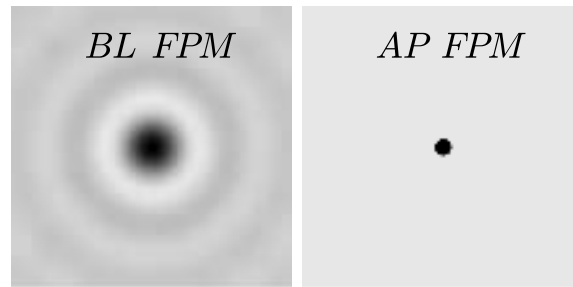}
\caption{The transmission of the focal plane masks of the band-limited (BL) and 
apodized pupil (AP) Lyot coronagraph designs shown in Fig.~\ref{fig:pupils}. The first 
zero of the BL mask transmission profile (a $1 - \jinc$ function) occurs at $8 \lambda/D$.  
The images shown are $64 \lambda/D$ across. This band-limited function is unphysical 
in that the transmission of the mask exceeds unity in some regions, although this detail 
does not complicate our analysis nor invalidate our conclusions. This mask induces a scale 
of $D/8$ in the Lyot pupil plane, which is seen in the residual spider diffraction shown in the 
panel (BL~b) of Fig.~\ref{fig:apblims}. The AP mask is $3.74 \lambda/D$ across.}
\label{fig:blcstop}
\end{figure*}

\begin{figure*}[htbp]
\epsscale{0.5}
\plotone{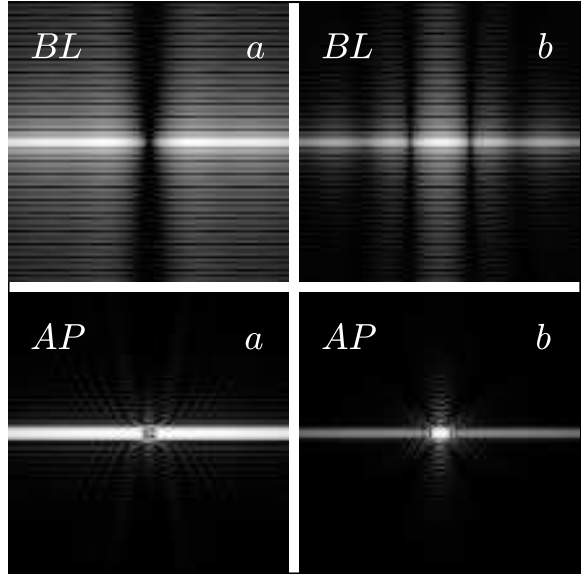}
\caption{Coronagraphic on-axis PSF of an apodized pupil (AP) and band-limited (BL) Lyot 
coronagraph in the presence of a single spider vane.  Without blocking the bright spider 
vanes in the Lyot pupil [case~(a)], the band-limited coronagraph shows better suppression 
of light from an on-axis star than the apodized pupil Lyot coronagraph.  When bright spider 
vanes in the Lyot plane are blocked  [case~(b)], the apodized pupil Lyot coronagraph performs 
better than the band-limited design.  The apodized design focal plane mask diameter is $3.74 
\lambda/D$, and the band-limited mask is the same as that described in  \ref{fig:blcstop}, with 
a scale size of $8\lambda/D$.  A logarithmic stretch between $1.0 \times 10^{-3}$ and  $1.0 
\times 10^{-8}$ is used for all these images.  Normalization is such that the peak intensity of 
the PSF without a focal plane mask, but through the optimized Lyot stop without spider 
obstructions is unity. }
\label{fig:apblims}
\end{figure*}

However, we should consider an appropriately optimized Lyot stop --- on a 
traditional or `classical' Lyot coronagraph (or a band-limited coronagraph) this stop 
is undersized relative to the entrance pupil.  In apodized pupil Lyot coronagraphs,
the Lyot stop is not undersized; it lets the entire graded entrance pupil through.
In the former case we would need to convolve $a_c(k)$ by the Fourier transform  
of this undersized pupil.  If the undersized classical Lyot stop is described by the 
function $A'(x)$, the classical or band-limited coronograph's ASF is
\begin{eqnarray} \label{ASFclassical} 
a_{c, LC} (k) &=& \epsilon a'(k) \conv \sinc(\epsilon k_1) \nonumber \\
& &   - \epsilon a'(k) \conv  \bigl{[} w(k)  \bigl{(} a(k) \conv \sinc(\epsilon k_1 )\bigr{)} \bigr{]}
\end{eqnarray}
(we note that $a'(k) \conv a(k) = a'(k)$ if the Lyot stop and entrance pupil are not 
apodized, and the Lyot stop support is a subset of the entrance pupil support).
\placefigure{fig:blcstop}
\placefigure{fig:apblims}

With a Lyot spider stop the first term disappears, leaving only the second term in
equation (\ref{ASFclassical}).  We see from this that the effects of spider diffraction 
are now concentrated behind the mask, as in the case of $a_c$, but because of a 
convolution with $a'(k)$, the diffracted light leaks out about a diffraction width outside 
the actual mask (in the final coronagraphic image plane).  Thus we can immediately 
conclude that for a hard-edged focal plane mask with diameter $s\lambda/D$, 
combined with an optimized Lyot spider stop, residual spider diffraction is very small 
outside a circle of diameter $s \lambda/D + \lambda/D'$ around the on-axis star ($D'$ 
being the Lyot stop outer diameter, projected back to the primary mirror).  Extending 
this argument to  obstructed apertures is straightforward: the spider diffraction will fall 
drastically at about a PSF equivalent width away from the geometric shadow of the 
focal plane mask in the final image plane.  The secondary obstruction will make for a 
larger distance scale for the fall-off of residual spider diffraction outside the focal 
plane mask's edge. For graded focal plane masks (such as Gaussian or band-limited 
coronagraphs), residual spider diffraction after using an optimized  Lyot spider stop 
extends everywhere that the mask shape function is non-zero.

Monochromatic coronagraphic images using a simple and a spider Lyot stop in a 
band-limited coronagraph are shown in Fig.~\ref{fig:apblims}, cases  {\it BL-a} and 
{\it BL-b} respectively. Here the perfect coronagraph on a clear unobstructed aperture
with no phase or amplitude errors produces no on-axis light whatsoever in the Lyot pupil 
after it is stopped down by the Lyot stop (Fig.~\ref{fig:pupils}, in the `no spider' column's 
Lyot stop intensity), or in the following coronagraphic image plane. A Lyot stop optimized 
without consideration of spider vanes lets through a horizontally-oriented `spider spike' of 
light (Fig.~\ref{fig:apblims}, {\it BL-a}). If there were two crossed spider vanes across the 
aperture, this image would show the usual cross observers associate with secondary 
support spiders. The `ringing' in the vertical direction has a period of the resolution 
element induced by the Lyot stop: approximately twice the size of the non-coronagraphic 
resolution $\lambda/D$ for this design. Removal of the bright strip of light from the Lyot
plane (where the geometrical image of the spider vane is located) by using a Lyot spider 
stop removes much of the light from the coronagraphic image (Fig.~\ref{fig:apblims}, 
{\it BL-b}). The dark vertical stripes are located at multiples of $8\lambda/D$ in this panel.
The focal plane mask structure of the band-limited coronagraph design in this example is 
seen by comparing this residual coronagraphic image with the focal plane mask itself 
(shown in Fig.~\ref{fig:blcstop}). 

For the apodized pupil Lyot coronagraph the convolution is with the Fourier transform of 
the full entrance pupil without apodization. This coronagraphic design does not undersize 
the Lyot pupil, so $D' = D$. For the unobscured aperture with a spider, this Fourier transform 
is the $\jinc$ function,  $2J_1(x)/x$ (where $J_1(x)$ is the Bessel function of the first kind, 
with index $1$) --- 
\begin{eqnarray} \label{ASFapod} 
a_{c, APLC} (k) &=& \epsilon\  \jinc(Dk) \conv a(k) \conv \sinc(\epsilon k_1) \nonumber \\
& &   - \epsilon\  \jinc(Dk) \conv  \bigl{[} w(k)  \bigl{(} a(k) \conv \sinc(\epsilon k_1 )\bigr{)} \bigr{]}.
\end{eqnarray}
We can treat centrally obscured apertures by using the difference of two $\jinc$ functions, 
although for simplicity we show the unobscured aperture case here.  Once again, an 
optimized Lyot spider stop will remove the first term in equation (\ref{ASFapod}), leaving 
the second term.  This term is reduced in field strength by a factor $\epsilon$ from what is 
essentially $A(0,0)$, and it is also restricted to a diffraction-width (or equivalent width of 
the PSF for obscured apertures) around the the projection of the focal plane mask on the 
final coronagraphic image plane.

\subsection{Apodized pupil and apodized occulter coronagraphs}

On a telescope with a secondary mirror obstructing the entrance aperture, it is possible to 
design Lyot coronagraphs such as apodized pupil Lyots, and Gaussian or band-limited 
coronagraphs.  The presence of spider vanes makes the apodized pupil design preferable 
to the apodized occulter designs. If we examine Figs.~\ref{fig:pupils} and \ref{fig:apblims}
in the case of the theoretically perfect band-limited design we can understand why this is 
the case.

The focal plane mask used in this example possesses a $1 - \jinc(\alpha k)$ transmission 
function.  $\alpha$ is chosen to produce a first zero in the $\jinc$ function at $k = 8\lambda/D$.
The width of the residual broad swath of low intensity light is twice the bandwidth or equivalent 
width of the function describing the focal plane mask (Fig.~\ref{fig:pupils}, Lyot plane on the left).

In the image plane of the band-limited coronagraph with the Lyot spider stop 
(Fig.~\ref{fig:apblims}   {\it BL-b}), we see that light diffracted by the spider vane spills into the 
coronagraphic image plane wherever the focal plane mask has a transmission that is not unity, 
\ie\ wherever our focal plane mask had any opacity whatsoever.  In fact the light occupies a 
slightly larger area (by about one resolution element) of the image after passage through an 
undersized Lyot stop.

With an apodized pupil design that utilizes a hard-edged focal plane mask, the on-axis 
coronagraphic image  (which is due almost entirely to spider vane diffraction when we use our 
analytical, circularly-symmetric apodizer) is localized to a circular area with a radius one 
resolution element larger than the original focal plane mask (Fig.~\ref{fig:apblims}   {\it AP-b}).
Thus, for the 8\ m Gemini telescope with an apodized pupil Lyot coronagraph with a 
$4\lambda/D$ diameter focal plane mask, all diffraction from the spider vanes is restricted to a 
disk $6\lambda/D$ in diameter.  Furthermore, the contrast ratio between this diffracted light
and a PSF taken with the same apodizer and Lyot stop, but no focal plane stop, is of the order 
of $10^{-6}$ with a simple analytical apodizer design.

\begin{figure*}[htbp]
\epsscale{.9}
\plotone{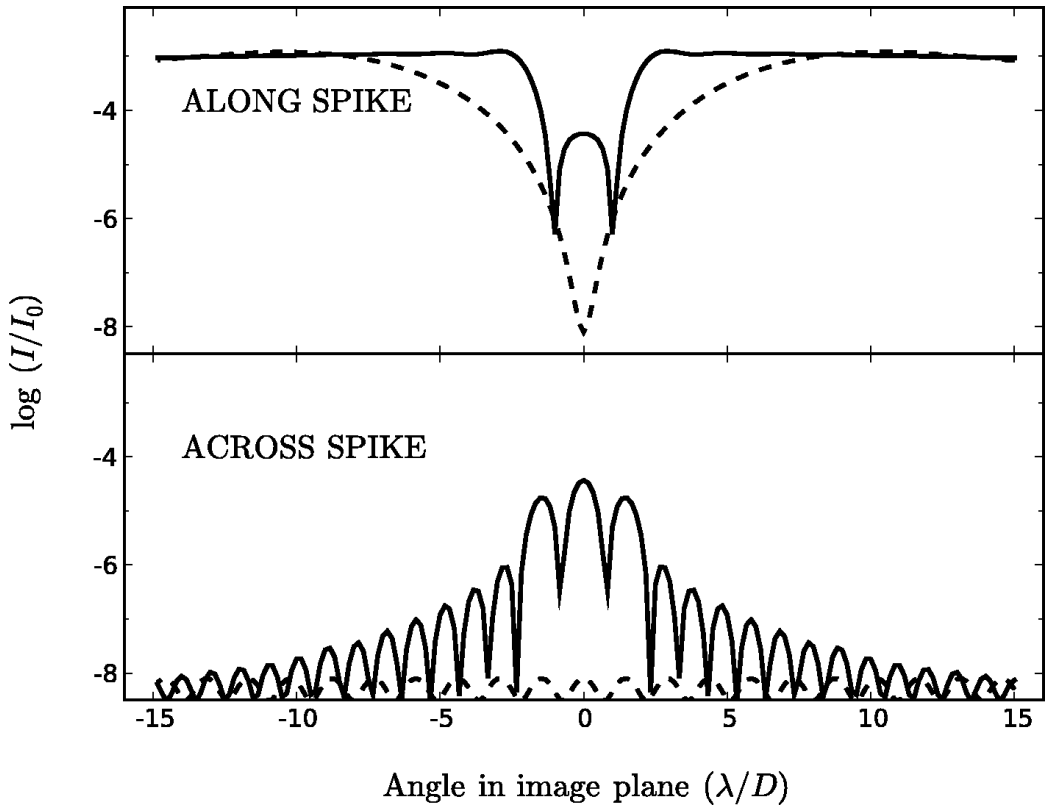}
\caption{Logarithmic intensity profiles of the coronagraphic PSF of the same designs shown
in Fig.~\ref{fig:apblims}.  Here the optimized Lyot stop does not block out the single spider vane.  
Diffracted light from the spider reduces contrast significantly. We utilize the same PSF 
normalization as in Fig.~\ref{fig:apblims}. The solid curve shows the apodized pupil Lyot 
coronagraphic PSF profile and the dashed line shows the band-limited coronagraphic PSF 
profile  (panel {\it AP-a} and {\it BL-a} in Fig.~\ref{fig:apblims}).  In this case coronagraphic 
performance is poor, although the band-limited design is less susceptible to diffracted light 
from the spider vane.}
\label{fig:spidcut}
\end{figure*}

\begin{figure*}[htbp]
\epsscale{.9}
\plotone{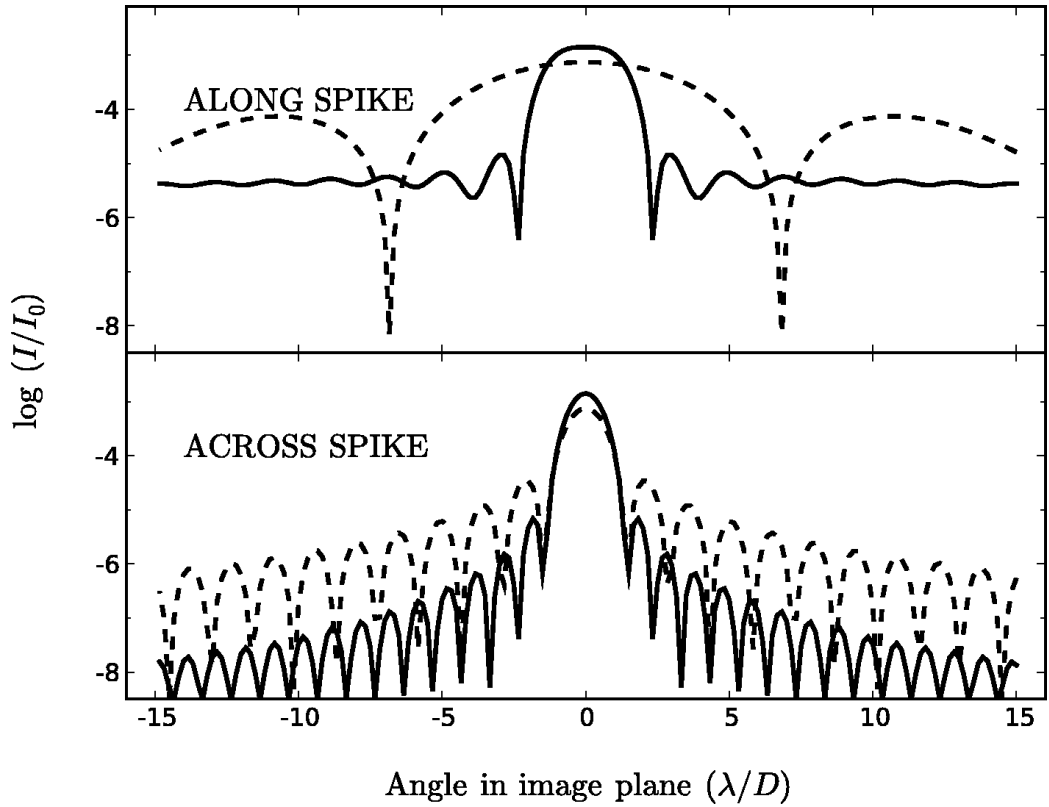}
\caption{Logarithmic intensity profiles of the coronagraphic PSF of the same designs shown
in Fig.~\ref{fig:apblims}.  In contrast with Fig.~\ref{fig:spidcut}, here the optimized Lyot stop 
blocks out the single spider vane. Diffracted light from the spider is thereby suppressed greatly.
We utilize the same PSF normalization as in Fig.~\ref{fig:apblims}. The solid curve shows the 
apodized pupil Lyot coronagraphic PSF profile,   and the dashed line shows the band-limited 
coronagraphic PSF profile (panel {\it AP-b} and {\it BL-b} in Fig.~\ref{fig:apblims}).  In this 
case coronagraphic performance is much improved, and the apodized pupil design contains 
the spread of the residual light diffracted by the spider far better than does the band-limited 
design.  The PSF of the apodized pupil Lyot coronagraph drops quickly with distance from the 
residual bright spider spike seen in panel {\it AP-b} of Fig.~\ref{fig:apblims}.}
\label{fig:spidstopcut}
\end{figure*}

\placefigure{fig:spidcut}
\placefigure{fig:spidstopcut}
This is visible in Fig.~\ref{fig:apblims}, where we see the structure of the focal plane mask in 
residual spider diffraction after using a Lyot spider stop. We simulated a single spider vane 
three pixels wide across a 170 pixel diameter aperture. We calculated the PSFs with the 
band-limited design described here, as well as with a numerical approximation of an 
analytical apodized pupil Lyot coronagraph optimized for a circular pupil \citep{SAF03}.  
While \citet{Soummer05} shows that it is possible to refine the apodized pupil coronagraph 
design to accomodate spiders, we used the circular aperture apodizer design because an 
azimuthally symmetric apodizer is easier to align in practice, and the analytical apodizing 
function is easy to generate numerically.  The particular example we present is taken from 
\citet{SAF03}. It has a 19\% throughput apodizer matched to a $3.74 \lambda/D$ focal plane 
mask diameter. Figs.~\ref{fig:spidcut} and \ref{fig:spidstopcut} show cuts through the PSFs 
of both designs, with and without Lyot spider stops. These cuts show that the apodized pupil 
coronagraphic design outperforms the band-limited design away from residual spider spikes
in the coronagraphic image when the bright spiders are blocked in the Lyot pupil, in spite of 
the fact that the apodized pupil design's focal plane mask equivalent width is smaller than 
that of our band-limited example. This localization of diffracted light due to spiders suggests 
that apodized pupil coronagraphic designs suit future ground-based adaptive optics 
coronagraphic instruments on existing telescopes.

\section {Discussion} 

We have demonstrated that even after masking out bright spiders in the Lyot plane of a Lyot 
coronagraph, some residual spider diffraction will be seen around the focal plane mask in the 
coronagraphic image. These effects cause a brightness proportional to the square of the spider 
thickness, but are localized to  a PSF-width around the focal plane mask.  Hard-edged masks 
show a stronger localization of the light diffracted by spiders than Gaussian or band-limited 
masks. This localization is desirable behavior for an extreme adaptive optics coronagraph on 
existing 8--10~m class telescopes, all of which possess secondary mirror support spider vanes.

The residual brightness due to spider vane diffraction will affect speckle statistics in this region 
\citep{AS04ApJL},  inducing a larger variance in intensity there. However, on hard-edged focal 
plane mask coronagraphs, masking out the bright spiders in the Lyot plane does enable good 
suppression of spider diffraction just one resolution element away from the mask edge. 
Opto-mechanical tolerances for on-axis telescope designs are looser than for extreme off-axis 
designs.  The advent of on-axis apodized pupil Lyot coronagraphs with good suppression of 
on-axis sources \citep{Soummer05} makes it important to understand the effects of spiders in 
Lyot coronagraphs and their modern variants when designing coronagraphs to search for 
extrasolar jovian companions using ground-based, next-generation adaptive optics systems on
today's 8--10~m class telescopes.  

\acknowledgements
We acknowledge frequent helpful discussions with R.~Soummer, and his generous contribution 
of a numerical realization of the apodized pupil Lyot coronagraph on an unobstructed circular 
aperture.   We are grateful to the referee for useful comments, and to P.~E.\ Hodge, P.~Greenfield, 
J.~T.\ Miller, and N.~Dencheva for their role in developing and supporting the Python Numarray 
module \citep{numarray02, pycon03}, wrapping the numerical Fourier transform library FFTW 
\citep{fftw} for Numarray, and providing support for matplotlib \citep{matplotlib}. We also thank the 
Space Telescope Science Institute's Research Programs Office and Director's Discretionary 
Research Fund for support. This work has also been supported by the National Science Foundation
Science and Technology Center for Adaptive Optics, managed by the University of California at 
Santa Cruz under cooperative agreement No.\ AST-9876783, and by the National Science 
Foundation under  Grant No.\ AST-0215793 and Grant No.\ AST-0334916.

\bibliographystyle{apj}
\bibliography{ms}

\begin{thebibliography}{27}
\expandafter\ifx\csname natexlab\endcsname\relax\def\natexlab#1{#1}\fi

\bibitem[{{Aime} \& {Soummer}(2004)}]{AS04ApJL}
{Aime}, C. \& {Soummer}, R. 2004, \apjl, 612, L85

\bibitem[{{Aime} {et~al.}(2001){Aime}, {Soummer}, \& {Ferrari}}]{Aime01aa}
{Aime}, C., {Soummer}, R., \& {Ferrari}, A. 2001, \aap, 379, 697

\bibitem[{{Aime} {et~al.}(2002){Aime}, {Soummer}, \& {Ferrari}}]{Aime02aa}
---. 2002, \aap, 389, 334

\bibitem[{{Burrows}(2005)}]{Bnat05}
{Burrows}, A. 2005, \nat, 433, 261

\bibitem[{{Burrows} {et~al.}(2004){Burrows}, {Sudarsky}, \& {Hubeny}}]{BSH04}
{Burrows}, A., {Sudarsky}, D., \& {Hubeny}, I. 2004, \apj, 609, 407

\bibitem[{{Digby} {et~al.}(2004){Digby}, {Oppenheimer}, {Newburgh}, {Brenner},
  {Shara}, {Mey}, {Mandeville}, {Makidon}, {Sivaramakrishnan}, {Soummer},
  {Graham}, {Kalas}, {Perrin}, {Roberts}, {Kuhn}, {Whitman}, \&
  {Lloyd}}]{DigbySPIE04}
{Digby}, A.~P., {Oppenheimer}, B.~R., {Newburgh}, L., {Brenner}, D., {Shara},
  M., {Mey}, J., {Mandeville}, C., {Makidon}, R.~B., {Sivaramakrishnan}, A.,
  {Soummer}, R., {Graham}, J.~R., {Kalas}, P., {Perrin}, M.~D., {Roberts},
  L.~C., {Kuhn}, J., {Whitman}, K., \& {Lloyd}, J.~P. 2004, in Proc. SPIE Vol.
  5490, Advances in Adaptive Optics, Roberto Ragazzoni and Domenico Bonaccini;
  Eds.

\bibitem[{Frigo \& Johnson(1997)}]{fftw}
Frigo, M. \& Johnson, S.~G. 1997, in Technical Report MIT-LCS-TR-728
  (Massachusetts Institute of Technology)

\bibitem[{Greenfield {et~al.}(2003)Greenfield, Miller, Hsu, \& White}]{pycon03}
Greenfield, P., Miller, J.~T., Hsu, J.-C., \& White, R.~L. 2003, in PyCon 2003
  Proceedings, ed. S.~Holden

\bibitem[{{Greenfield} {et~al.}(2002){Greenfield}, {Miller}, {Hsu}, \&
  {White}}]{numarray02}
{Greenfield}, P., {Miller}, T., {Hsu}, J.-C., \& {White}, R.~L. 2002, in ASP
  Conf. Ser. 281: Astronomical Data Analysis Software and Systems XI, 140--+

\bibitem[{{Guyon}(2003)}]{Guyon03}
{Guyon}, O. 2003, \aap, 404, 379

\bibitem[{Hunter(2005)}]{matplotlib}
Hunter, J. 2005, The Matplotlib User's Guide (University of Chicago Medical
  School)

\bibitem[{{Kasdin} {et~al.}(2003){Kasdin}, {Vanderbei}, {Spergel}, \&
  {Littman}}]{Kasdin03}
{Kasdin}, N.~J., {Vanderbei}, R.~J., {Spergel}, D.~N., \& {Littman}, M.~G.
  2003, \apj, 582, 1147

\bibitem[{{Kuchner} \& {Traub}(2002)}]{KT02}
{Kuchner}, M.~J. \& {Traub}, W.~A. 2002, \apj, 570, 900

\bibitem[{{Lloyd} {et~al.}(2003){Lloyd}, {Gavel}, {Graham}, {Hodge},
  {Sivaramakrishnan}, \& {Voit}}]{LGG03}
{Lloyd}, J.~P., {Gavel}, D.~T., {Graham}, J.~R., {Hodge}, P.~E.,
  {Sivaramakrishnan}, A., \& {Voit}, G.~M. 2003, in High-Contrast Imaging for
  Exo-Planet Detection. Edited by Alfred B. Schultz. Proceedings of the SPIE,
  Volume 4860, pp. 171-181 (2003)., 171--181

\bibitem[{{Lloyd} \& {Sivaramakrishnan}(2005)}]{Lloyd05}
{Lloyd}, J.~P. \& {Sivaramakrishnan}, A. 2005, \apj, 621, 1153

\bibitem[{{Lyot}(1939)}]{Lyot39}
{Lyot}, B. 1939, \mnras, 99, 580

\bibitem[{{Makidon} {et~al.}(2005){Makidon}, {Sivaramakrishnan}, {Perrin},
  i~{Roberts}, {Oppenheimer}, {Soummer, R.}, \& {Graham}}]{MSP05}
{Makidon}, R.~B., {Sivaramakrishnan}, A., {Perrin}, M.~D., i~{Roberts}, L.~C.,
  {Oppenheimer}, B.~R., {Soummer, R.}, \& {Graham}, J.~R. 2005, \pasp, {in
  press}

\bibitem[{{Oppenheimer} {et~al.}(2004){Oppenheimer}, {Digby}, {Newburgh},
  {Brenner}, {Shara}, {Mey}, {Mandeville}, {Makidon}, {Sivaramakrishnan},
  {Soummer}, {Graham}, {Kalas}, {Perrin}, {Roberts}, {Kuhn}, {Whitman}, \&
  {Lloyd}}]{OppenheimerSPIE04}
{Oppenheimer}, B.~R., {Digby}, A.~P., {Newburgh}, L., {Brenner}, D., {Shara},
  M., {Mey}, J., {Mandeville}, C., {Makidon}, R.~B., {Sivaramakrishnan}, A.,
  {Soummer}, R., {Graham}, J.~R., {Kalas}, P., {Perrin}, M.~D., {Roberts},
  L.~C., {Kuhn}, J., {Whitman}, K., \& {Lloyd}, J.~P. 2004, in Proc. SPIE Vol.
  5490, Advances in Adaptive Optics, Roberto Ragazzoni and Domenico Bonaccini;
  Eds.

\bibitem[{{Oppenheimer} {et~al.}(2003){Oppenheimer}, {Sivaramakrishnan}, \&
  {Makidon}}]{OSM03}
{Oppenheimer}, B.~R., {Sivaramakrishnan}, A., \& {Makidon}, R.~B. 2003,
  {Imaging Exoplanets: The Role of Small Telescopes} (The Future of Small
  Telescopes In The New Millennium.~Volume III - Science in the Shadow of
  Giants), 155

\bibitem[{{Roberts} \& {Neyman}(2002)}]{LRCN02}
{Roberts}, L.~C. \& {Neyman}, C.~R. 2002, \pasp, 114, 1260

\bibitem[{{Rouan} {et~al.}(2000){Rouan}, {Riaud}, {Boccaletti}, {Cl{\' e}net},
  \& {Labeyrie}}]{RRB00}
{Rouan}, D., {Riaud}, P., {Boccaletti}, A., {Cl{\' e}net}, Y., \& {Labeyrie},
  A. 2000, \pasp, 112, 1479

\bibitem[{{Sivaramakrishnan} {et~al.}(2001){Sivaramakrishnan}, {Koresko},
  {Makidon}, {Berkefeld}, \& {Kuchner}}]{Sivaramakrishnan01}
{Sivaramakrishnan}, A., {Koresko}, C.~D., {Makidon}, R.~B., {Berkefeld}, T., \&
  {Kuchner}, M.~J. 2001, \apj, 552, 397

\bibitem[{{Sivaramakrishnan} {et~al.}(2005){Sivaramakrishnan}, {Soummer},
  Sivaramakrishnan, Lloyd, Oppenheimer, Perrin, \&
  Makidon}]{Sivaramakrishnan05blc}
{Sivaramakrishnan}, A., {Soummer}, R., Sivaramakrishnan, A.~V., Lloyd, J.~P.,
  Oppenheimer, B.~R., Perrin, M.~D., \& Makidon, R.~B. 2005, \apj, submitted

\bibitem[{{Sivaramakrishnan} \& {Yaitskova}(2005)}]{SY05}
{Sivaramakrishnan}, A. \& {Yaitskova}, N. 2005, \apjl, 626, L65

\bibitem[{{Soummer}(2005)}]{Soummer05}
{Soummer}, R. 2005, \apjl, 618, L161

\bibitem[{{Soummer} {et~al.}(2003{\natexlab{a}}){Soummer}, {Aime}, \&
  {Falloon}}]{SAF03}
{Soummer}, R., {Aime}, C., \& {Falloon}, P.~E. 2003{\natexlab{a}}, \aap, 397,
  1161

\bibitem[{{Soummer} {et~al.}(2003{\natexlab{b}}){Soummer}, {Dohlen}, \&
  {Aime}}]{SDA03}
{Soummer}, R., {Dohlen}, K., \& {Aime}, C. 2003{\natexlab{b}}, A\&A, 403, 369

\end{thebibliography}
\newpage

\end{document}